\documentclass[twocolumn,showpacs,prb,nobibnotes,superscriptaddress]{revtex4}

\usepackage{color}
\usepackage{dcolumn}
\usepackage{bm}
\usepackage{amsmath}
\usepackage{graphicx}
\graphicspath{{./figures/}}
\usepackage{gensymb}
\newcommand{\angstrom}{$\,$\AA{}}

\newcommand{\SNW}{Sr$_{2}$NiWO$_{6}$}

\newcommand{\TNeel}{$T_N$}

\usepackage{soul}
\setstcolor{red}
\setul{}{1.5pt}

\usepackage{multirow,varwidth}
\newcommand{\multilineC}[1]{\begin{tabular}[b]{@{}c@{}}#1\end{tabular}}

\begin{document}

\title{Ab initio investigation of magnetic ordering in the double perovskite  \SNW}

\author{Nafise Rezaei}
\email{n-rezaee@ph.iut.ac.ir}
\affiliation{Department of Physics, Isfahan University of Technology, Isfahan 84156-83111, Iran}
\author{Tayebehsadat Hashemifar}
\affiliation{Department of Physics, Isfahan University of Technology, Isfahan 84156-83111, Iran}
\affiliation{Department of Physics, Faculty of Science, Shahrekord University, 
             Shahrekord 88186-34141, Iran}
\thanks{The first two authors have equal contributions to this work}
\author{Mojtaba Alaei}
\affiliation{Department of Physics, Isfahan University of Technology, Isfahan 84156-83111, Iran}
\author{Farhad Shahbazi}
\affiliation{Department of Physics, Isfahan University of Technology, Isfahan 84156-83111, Iran}
\author{S. Javad Hashemifar}
\affiliation{Department of Physics, Isfahan University of Technology, Isfahan 84156-83111, Iran}
\author{Hadi Akbarzadeh}
\affiliation{Department of Physics, Isfahan University of Technology, Isfahan 84156-83111, Iran}

\date{\today}

\begin{abstract}

{\it Ab initio} calculations, GGA/GGA+$U$, are used to propose a spin Hamiltonian 
for the B-site ordered double perovskite, \SNW{}.
Our results show that the exchange interaction constants between the next nearest neighbors in 
both intra- and inter- $ab$ plane ($J_2$ and $J_{2c}$) 
are an order of magnitude larger than the ones between the nearest neighbors ($J_1$ and $J_{1c}$).
Employing the Monte Carlo simulation, we show that the obtained Hamiltonian properly describes 
the finite temperature properties of \SNW{}.
Our {\em ab initio} calculations also reveal a small  magnetic anisotropy and  non-trivial bi-quadratic interaction 
between the nearest inter-$ab$ plane neighbors, which play essential roles 
in stabilizing the type-II anti-ferromagnetic ground state of \SNW{}. 

\end{abstract}


\maketitle

\section{Introduction}

Ordered double perovskites with general chemical formula A$_2$BB$^\prime$O$_6$, 
have received great attention owing to the magnetic interactions tunable by substitutions 
on B and $\mathrm{B'}$ ions.\cite{vasala-rev,howard,Hossain}
For instance the substitution of Mo by W in Sr$_2$CuMoO$_6$, tends to decreasing the Curie-Weiss 
temperature from -116 to -300\,K.\cite{vasala2014}
The wide variety of $\mathrm{A_2BB'O_6}$ compounds with different A, B, and $\mathrm{B'}$ 
ions represents various novel properties such as colossal magnetoresistance 
in $\mathrm{Sr_2FeMoO_6}$,\cite{Kobayashi1998}
half-metallicity in $\mathrm{Sr_2CrWO_6}$,\cite{philipp2003}
multiferroics in $\mathrm{Sr_2NiMoO_6}$, \cite{Hsu2012, Kumar2010, prasat2011}
and Pb$_2$FeMeO$_6$ (Me=Nb, Ta, Sb) ceramics~\cite{Gusev,Laguta},
photovoltaics in Bi$_2$FeCrO$_6$  \cite{Nechache} and Sc$_2$FeCrO$_6$ ~\cite{Cai}  
and low dimensional anti-ferromagnetic (AFM) behavior in $\mathrm{Ba_2CuB'O_6}$ and 
$\mathrm{Sr_2CuB'O_6}$~($\mathrm{B'=W,Te}$) compounds.\cite{iwanaga-1999}

In the majority of  magnetic ordered double 
perovskites A$_2$BB$^\prime$O$_6$, 
 B-site, B$'$-site or both could be occupied by transition metal magnetic ions.
For the cases that B is magnetic and B$'$ is a diamagnetic ion, the magnetic ions interact
to each other through B-O-B$'$-O-B bonds~(Fig.\ref{fig:snwo}). 
The magnetic B ions  can interact  to each other through the direct and super-exchange interactions.
Due to the large distance between the B ions the direct exchange interaction is negligible, 
hence the dominant magnetic interaction 
would be the super-exchange interaction mediated by the B$'$ and O ions. 
The B-B$'$-B angle is $90\degree$ for the nearest and 180\degree\ for 
the next nearest neighbors~(Fig.\ref{fig:snwo}),
which could make the nearest neighbor (NN) super-exchange interaction much smaller than 
the next nearest neighbor (NNN) interaction. 
The dependence of super-exchange interaction on the bond angle 
is given by the Goodenough-Kanamori-Anderson rules~\cite{Goodenough,Kanamori},
according which the super-exchange is AFM and its strength is maximum for $180\degree$ bond angle.

Typically, ordered double perovskites show low temperature AFM ordering,
except for some compounds including $\mathrm{La_2BMnO_6~(B=Mg,Co,Ni,Cu)}$ group
which represent ferromagnetic (FM) ordering.\cite{blasse-ferro}

\SNW{} is an B-site ordered double perovskite in which the magnetic Ni$^{+2}$ ion resides on 
the B site and hexavalent diamagnetic W$^{+6}$ ion occupies the B$'$ locations. 
Its crystalline structure at room temperature is tetragonal and transforms to 
the cubic symmetry above 520\,K.
The lattice distortion tilts the Ni-O-W angle from 180\degree\ to 165\degree\ 
in the $ab$ plane.\cite{iwanaga}
\SNW{} exhibits a sharp transition to a type-II AFM (AFM-II) 
spin ordering below the N\'eel temperature 54\,K.~\cite{iwanaga, Blum2015}
Analysis of the spin-wave excitation spectrum indicates that the 90\degree\ super-exchange 
interaction in \SNW\ is much smaller than the $180\degree$ one.\cite{todate1999,todate1995}
On the contrary, Iwanaga {\it et al.} argued that these
magnetic interactions in \SNW\ are comparable.\cite{iwanaga}
Hence, the relative strength of these super-exchange interactions in \SNW\ 
is a matter of dispute. 
Furthermore, the existence of a sharp peak in the magnetic susceptibility of \SNW{}, 
unlike $\mathrm{Sr_2CuWO_6}$, is an indication of a three dimensional ordering in this compound.  

The frustration of the exchange interactions between the spins could lead to  magnetic degeneracy
in anti-ferromagnetic materials.
The {\it{fcc}} magnetic lattice with the anti-ferromagnetic NN and NNN interactions is an example of 
the frustrated magnets. This lattice is composed of four Heisenberg anti-ferromagnetic cubic sub-lattices, in a way that
the sum of first neighboring ion magnetic fields at a given site vanishes. 
This results to four independent magnetic sub-lattices with AFM ordering. The magnetic moment directions of the sub-lattices
are not constrained on each other.
Such a freedom to select the relative magnetic moment direction can be lifted by including some higher order exchange interactions such as bi-quadratic interaction or single-ion interaction generated by the spin-orbit effect.
Neglecting the small tetragonal distortion in ordered double perovskites like \SNW,
 the magnetic lattice turns to be {\it{fcc}}, hence one expects the emergence of frustration in these compounds.
In this work, we study \SNW{} as a prototype of a (rock-salt) ordered double perovskite to shed light
on the magnetic features of these  compounds.
Moreover, the experimental spin-wave excitation spectrum obtained for \SNW{}~\cite{todate1999,todate1995}
will help us to compare the parameters of our spin Hamiltonian with those extracted by fitting the spin-wave spectrum 
using linear spin wave theory.

In this study,
we employ density functional calculations including Hubbard correction 
to build a spin model Hamiltonian for \SNW.
We show that the NNN exchange couplings between the intra- and inter- $ab$ plane Ni ions
are in the same order, 
which guarantees the three-dimensional magnetic ordering in \SNW.
We also discuss the thermodynamic properties of the obtained model Hamiltonian 
using classical Monte Carlo (MC) simulation.
In addition to the Heisenberg exchange couplings, we consider the bi-quadratic 
magnetic interaction and magnetic anisotropy interaction in the spin Hamiltonian and argue the key role 
of these interactions in finding the correct magnetic ground state of \SNW.
To our knowledge, there is no {\em ab initio} paper which considers the bi-quadratic interaction in these materials. 

The paper is organized as the following. Section~\ref{method} discusses 
the {\em ab initio} methods to construct the spin Hamiltonian and also the details of MC simulation. 
The results and discussions are given in section~\ref{results} and 
section~\ref{conclusions} is devoted to the conclusions.

\section{Method}
\label{method}

\begin{figure}[tb!]
\begin{center}
\includegraphics[width=0.45\textwidth,angle= 0.00,clip]{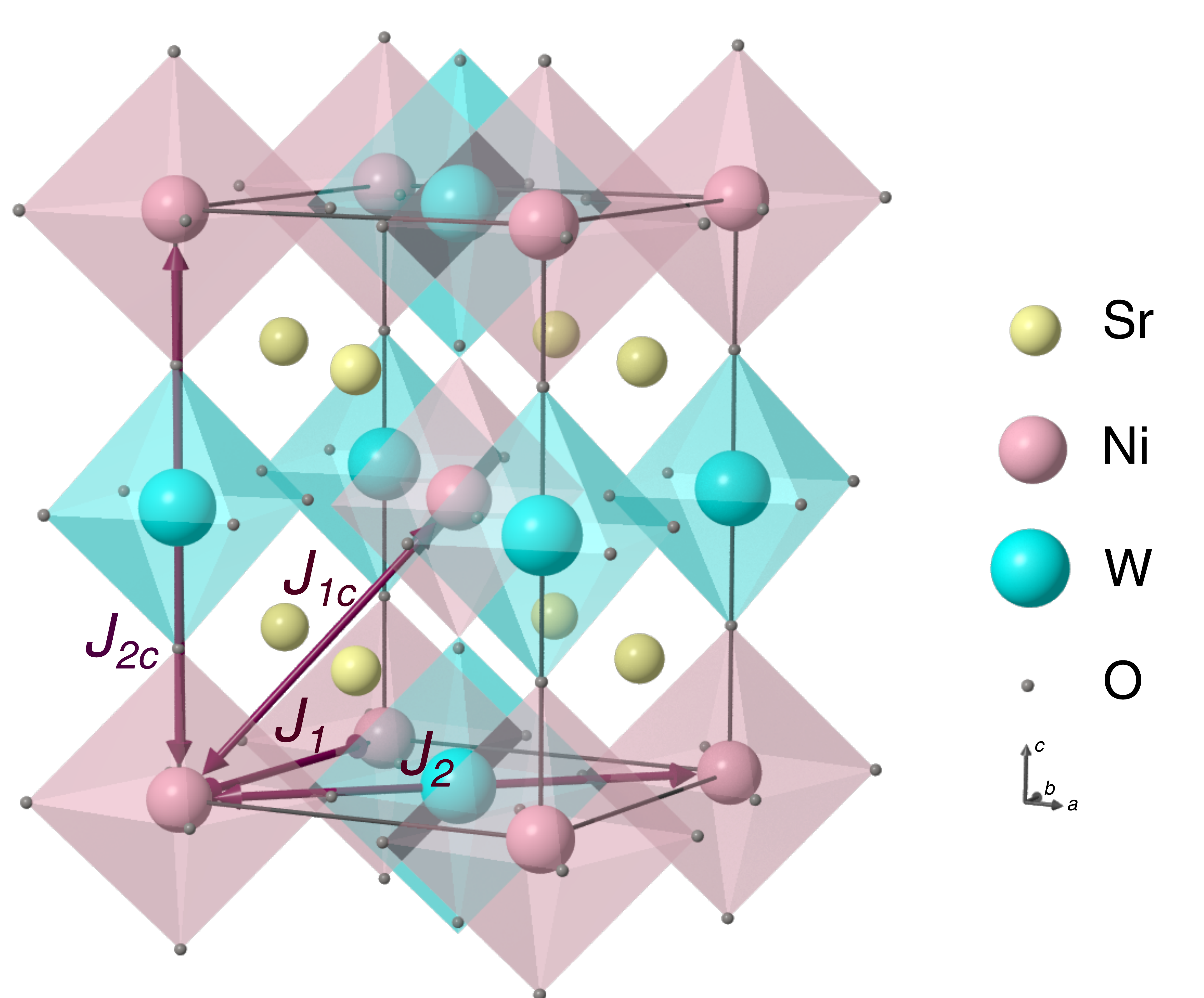}
\end{center}
\caption{(Color online) 
          Crystal structure of \SNW. The thick violet arrows show 
          the nearest and next nearest neighbors at the intra-$ab$ plane ($J_1$ and $J_2$) 
          and inter-$ab$ plane ($J_{1c}$ and $J_{2c}$), respectively.\cite{povray}}
\label{fig:snwo}
\end{figure}

The major part of {\em ab initio} calculations were done by 
the Quantum ESPRESSO (QE) package,\cite{Giannozi2009} 
which is based on density functional theory (DFT).
To treat electron-nucleus interaction, the projector-augmented wave (PAW) 
pseudo-potentials were employed.
The exchange-correlation potential was approximated by the Perdew-Burke-Ernzerhof (PBE)
functional within the generalized gradient approximation (GGA).\cite{Perdew1996}
To improve the on-site Coulomb repulsion of the localized d electrons, 
we have applied the GGA+$U$ method in a simplified approach by Dudarev,\cite{SLDAU} 
which only needs an effective Hubbard parameter ($U_{\rm{eff}}$).
We used $8\times8\times6$ k-point meshes for Brillouin zone sampling of
the primitive unit cell (which contains two formula units). 
The experimental crystal structure  was taken from Ref.~[\onlinecite{iwanaga}].
An energy cutoff of 40\,Ry (440\,Ry) was chosen for the wave function (electron density) 
expansion in the plane wave basis set.
Higher energy cutoffs were chosen for the lattice and site geometry optimization
(50 and 550\,Ry for wave function and density expansion, respectively).
We have estimated the $U_{\rm{eff}}$ parameter by using the linear response (LR) method.\cite{Cococcioni2005}
For these calculations, a $2\times 2\times 2$ supercell, containing 16 Ni atoms, was used.
We employed the full-potential linearized augmented plane wave (LAPW) method,
using Fleur code~\cite{fleur}, to verify PAW  pseudo-potentials.
For LAPW calculation, we set $k_{\rm max}=4.5$ a.u.$^{-1}$, and we chose  2.0, 2.0,
2.2  and 1.4 a.u. for muffin-tin radius of  Sr, Ni, W and O, respectively.

To find an effective spin Hamiltonian, the collinear spin-polarized DFT results 
were mapped to the Heisenberg Hamiltonian given by:
\begin{equation}
H= -\frac{1}{2}\sum_{i,j} J_{ij} {\hat{\bf n}}_i \cdot {\hat{\bf n}}_j
\end{equation}
where $\hat{\bf n}_i$ denotes an unit vector in the direction of 
the magnetic moment at the $i$-th lattice site, 
and $J_{ij}$'s are Heisenberg exchange constants describing the strength of magnetic coupling 
between the magnetic ions residing on the $i$-th and $j$-th sites. 
To derive the exchange constants, the DFT total energy of various magnetic configurations 
were calculated.
Then by employing the least-square method,
the NN and NNN exchange coupling at the intra-$ab$ plane ($J_1,~\mathrm{and}~J_2$) and 
inter-$ab$ plane ($J_{1c},~\mathrm{and}~J_{2c}$) were computed (Fig.~\ref{fig:snwo}).
\begin{figure*}[tb!]
\begin{center}
\includegraphics[width=1.00\textwidth,angle= 0.00,clip]{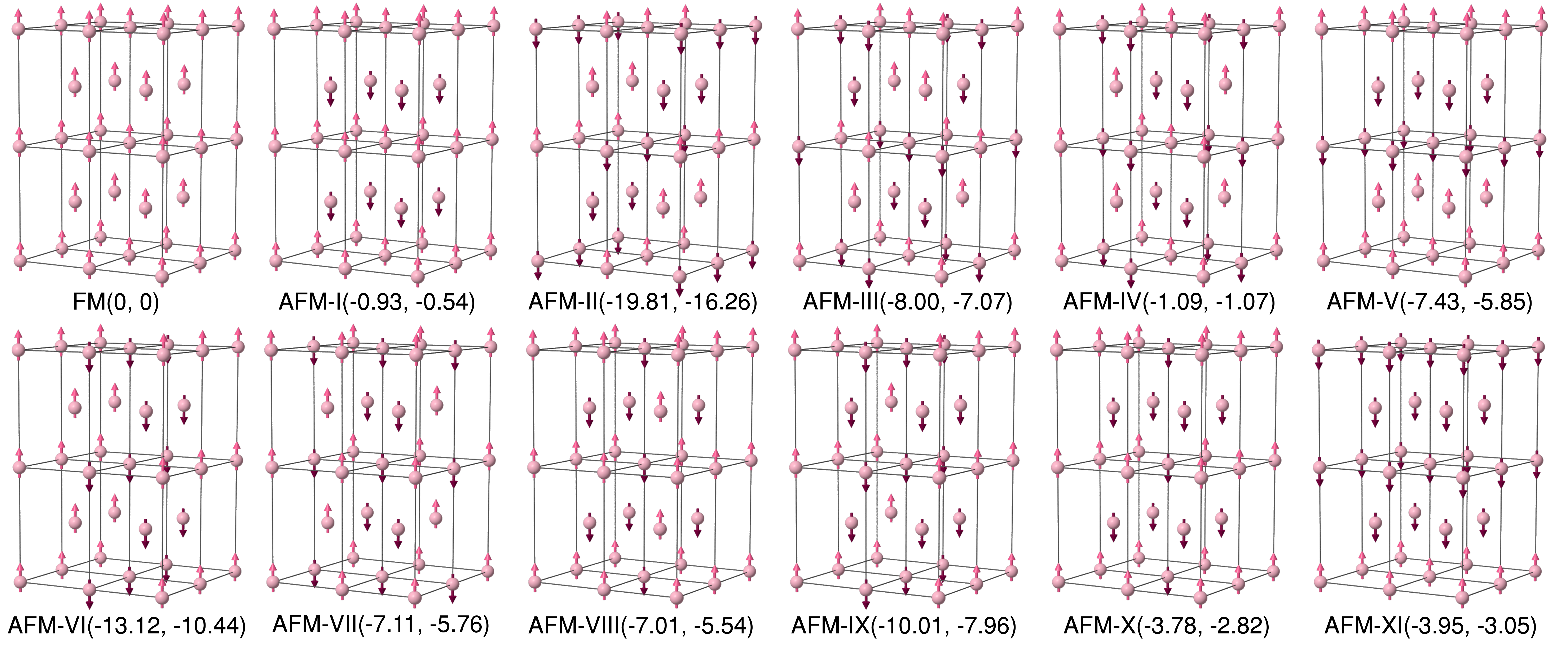}
\end{center}
\caption{(Color online) 
  Schematic representation of the Ni spin moments in the various 
  magnetic configurations used for calculating the exchange constants.
  The numbers in parentheses are the total energy difference of each magnetic 
  configuration (meV/f.u.) respect to the ferromagnetic (FM) configuration for $U_{\rm{eff}}$ = 5 and 4.72\,eV, respectively.}
\label{fig:st}
\end{figure*}


For the Ni ion with $S=1$, a bi-quadratic interaction 
$\sum _{i>j} B_{ij} ({\hat{\bf n}}_i \cdot {\hat{\bf n}}_j)^2$ is also expected.~\cite{Fazekas}  
To estimate the bi-quadratic couplings $B_{ij}$,
we used LAPW Fleur code~\cite{fleur} which is more specialized for non-collinear spin-polarized DFT.
Because of the existence of heavy element W in  \SNW, we  investigated the
effect of the magnetic anisotropy, $\Delta \sum _{i}  ({\hat{\bf n}}_i \cdot {\hat{\bf z}})^2$, where $\Delta$ denotes the strength of anisotropy. 
These calculations also were done within LAPW method  by including the spin-orbit coupling (SOC) and Hubbard correction (GGA+$U$+SOC).

In the end, we have used classical MC simulations to investigate 
the finite temperature properties of the obtained spin Hamiltonian.
The parallel tempering MC method was carried on a lattice size $N=2\times12^3$, 
and the uniform temperature range including 64 temperatures was selected.
We used $1\times10^6$ MC steps per spin for equilibration and  $1\times 10^6$ MC steps for sampling.
 To reduce the correlation between the data, we skipped 10 MC steps between the data collections. 
In parallel tempering algorithm, we allowed the spin configurations at 
the different temperatures to swap with each other after 10 MC steps.
\begin{table*}[tb!]
\caption{\label{tab:j}
  Obtained exchange constants (meV) and N\'eel temperature \TNeel (K)
  at different values of the Hubbard parameter (eV) for experimental and 
  optimized crystal structures of \SNW within PAW and LAPW method.
  The calculations (PAW/GGA$+U$) with $U_{\rm{eff}}=0, 5, 6, 7$\,eV were done in the experimental structure 
  and the one with $U_{\rm{eff}}=4.72$\,eV corresponds to the optimized structure.
  All of LAPW calculations have been done by using experimental structure.  
  The last row indicate INS results for exchange parameters~\cite{todate1999}. 
  For each exchange parameters set (except INS results), \TNeel{} is derived from MC simulations. 
}
\begin{ruledtabular}
\begin{tabular}{c|ccccccc}
 \vspace{-3mm}  \\  \vspace{1mm}
  & $U_{\rm{eff}}$ (eV) &  $J_1$ (meV)  & $J_{1c}$ (meV)   &  $J_2$ (meV)  &  $J_{2c}$ (meV)  & \TNeel (K)    \\
\hline
\multirow{5}{*}{PAW}
  & 0              & $-0.36$ &  $-0.34$   & $-8.06$ &  $-9.07$   & $139$ \\
  & 5              & $-0.16$ &  $-0.12$   & $-3.06$ &  $-3.45$   & $ 52$ \\
  & 6              & $-0.14$ &  $-0.10$   & $-2.52$ &  $-2.83$   & $ 43$ \\
  & 7              & $-0.11$ &  $-0.08$   & $-2.05$ &  $-2.30$   & $ 35$ \\   
  & 4.72           & $-0.24$ &  $-0.05$   & $-2.44$ &  $-2.90$   & $ 43$ \\  \hline
\multirow{2}{*}{LAPW}
  & 0             & $-0.54$  &  $-0.47$    & $-8.22$  & $-10.18$  & $146$ \\
  & 5 ($U=6.0$, $J_{H}$=1.0)             & $-0.27$  &  $-0.18$   & $-1.87$  & $-2.90$   &  $36$ \\ \hline
\multirow{2}{*} {\multilineC{Exp. \\ (INS)~\cite{todate1999}}}
&               &                   &  &                      &  &         \\ 
&               &  $-0.02\pm 0.08$      &  &  $-1.81\pm 0.09$         &  &  $54$~\cite{iwanaga, Blum2015}   \\
\end{tabular}
\end{ruledtabular}
\end{table*}

\section{Results and Discussion}
\label{results}

\subsection{Spin Hamiltonian}

\SNW{} crystallizes in the tetragonal space group $I4/m$ with cell parameters 
(a=5.5571, c=7.9133 \angstrom ).\cite{iwanaga}
As shown in Fig.~\ref{fig:snwo}, the transition metal ions including the magnetic 
Ni$^{+2}$ ions and non-magnetic W$^{+6}$ ions are located 
at the center of the oxygen octahedra.
In order to have an insight into the DFT magnetic ground state, different magnetic configurations
are considered, as presented in Fig.~\ref{fig:st}.
The calculations are performed within GGA, and GGA+$U$ approximations for 
the experimentally-identified as well as {\em ab initio} optimized crystal structures.
For any on-site Hubbard parameter $U_{eff}$ varying from 0 to 7\,eV, 
the magnetic ground state of \SNW{} is the AFM-II ordering (Fig.\ref{fig:st}), 
consistent with the experimental observation.\cite{todate1999,todate1995}
In AFM-II, each Ni ion aligns its moment parallel to the half and 
anti-parallel to the other half of its nearest neighbors, whose number is  $4$ in 
$ab$ plane and $8$ out of  this plane. 
However, for both the intra- and inter- $ab$ plane NNN the direction of magnetic moments 
are anti-parallel (Fig.\ref{fig:st}).  

To evaluate the on-site coulomb repulsion $U_{\rm{eff}}$, 
we use the linear response (LR) method.\cite{Cococcioni2005}
In LR approach, a perturbed repulsive coulomb interaction is applied as
a small shift of potential on $d$ levels such that the response of system 
to this perturbation remains linear. 
Using the experimental structure of \SNW{}, the $U_{\rm {eff}}$ converges 
to 6.2\,eV for Ni, independent of magnetic ordering. 
Therefore, in the GGA+$U$ calculations with the experimental structure we take 
the values 5, 6, and 7\,eV for the Hubbard parameter.  

To obtain consistent results in an {\em ab initio} theory, 
one needs to include all relevant details such as optimized structural geometry. 
Therefore,  using the GGA+$U$, we optimize structural geometry. 
For a fully consistent result, we also estimate $U_{\rm {eff}}$ in a self-consistent LR  (SCLR) scheme~\cite{U-SC}. 
For this purpose, in each step, the crystal structure are optimized in the GGA+$U$ calculation 
with the value of $U$ parameter obtained from the previous step.
Given the new structure, the value of $U$ is updated in the SCLR scheme.
Iterating this process makes the value of $U$ to converge to a constant.
Starting from $U_{\rm{eff}}=6.2$\,eV of the experimental structure, 
we find that the on-site Hubbard parameter converge to 4.72\,eV.
The total energy difference of the considered magnetic configurations
 and the FM state are reported in Fig.~\ref{fig:st} for the experimental structure and {\em ab initio} optimized structure with  $U_{\rm{eff}}=5$\,eV and
 $U_{\rm{eff}}=4.72$\,eV, respectively.

Now, we proceed to find the spin Hamiltonian. 
For this purpose, we map the resulting total energies onto the Heisenberg Hamiltonian. 
The relevant exchange constants ($J_1, J_{1c}, J_{2}, J_{2c}$), for the experimental
structure with $U_{\rm {eff}}=0, 5, 6, 7$\,eV and the optimized structure 
with $U_{\rm {eff}}=4.72$\,eV are listed in table~\ref{tab:j}, showing that 
all the couplings are AFM. The details of this calculation are given in Appendix A.   
The small difference between the energy of AFM-I and FM magnetic configurations 
(see Fig.~\ref{fig:st}) justifies the smallness of the inter-$ab$ 
plane NN coupling $J_{1c}$. 
Indeed one can simply find $J_{1c}=\mathrm{(E_{AF-I}-E_{FM})/8}$.

Table~\ref{tab:j} also shows that NNN coupling constants ($J_{2}, J_{2c}$) 
are an order of magnitude larger than the NN ones ($J_1, J_{1c}$).  
Indeed the 90\degree\ Ni-W-Ni bond angle in both intra- and inter- $ab$ planes makes 
the super-exchange interaction between the NN magnetic ions too weak.
On the other hand, the Ni-W-Ni angle for both the intra- and inter- $ab$ plane NNN ions 
is 180\degree\ which substantially enhances the NNN exchange constants.
Moreover, the Ni-O-W bond angle in the intra-$ab$ plane (165.8\degree) is slightly 
smaller than the corresponding bond angle in the inter-$ab$ plane (180\degree), 
which somewhat enhances $J_{2c}$ compared with $J_{2}$.  
It can be seen from table~\ref{tab:j} that the coupling constants decrease by 
increasing $U_{\rm{eff}}$, which is a consequence of attenuating the hopping amplitude 
of neighboring $d$ electrons as the expense of enlarging the on-site Coulomb repulsion. 

To check how the exchange constants depend on the method, 
we employ the LAPW method and compare its results to those obtained by PAW.
For GGA (i.e. $U=0$) there is 10$\%$ (in average) discrepancy in exchange constants between the two methods which is reasonable.
We repeat the LAPW calculation using GGA+$U$.
It is worthy to mention that GGA+$U$ implementation in Fleur LAWP code
is based on  Liechtenstein's approach~\cite{LDAU} which includes two parameters; $U$ as on-site
Coulomb repulsion and $J_H$ as on-site (Hund) exchange. 
However, in the PAW/GGA+$U$ method we employ the Dudarev's approach~\cite{SLDAU} which uses only one parameter i.e. $U_{\rm {eff}}$. 
Generally, the relationship between $U_{\rm {eff}}$ , $U$ and $J_H$ is $U_{\rm {eff}} = U-J_H$.
Knowing that in many oxides $J_H \sim 1\,$~eV~\cite{Anisimov1991,Vaugier}, in this work we set $J_H$ to 1\,eV.
In principle, one has to calculate $U$ and $J_H$                     
in LAPW/GGA+$U$ but as a rough approximation, we use $U=U_{\rm {eff}}+J_H$ where $U_{\rm {eff}}$ is the PAW value. 
The exchange parameters obtained by LAPW/GGA+U ($U=6.0$, $J_H=1.0$\, eV) are reported in Table I.
These results are comparable with those obtained by PAW when $U_{\rm {eff}}=5-6\,$ eV is used. 
 

Linear spin-wave (LSW)  fitting of excitation spectrum obtained by inelastic neutron scattering (INS)  experiment, 
results in $J_1 \approx -0.02$\,meV  and $J_2 \approx -1.81$\,meV.~\cite{todate1999,todate1995}
The discrepancy between our result and LSW comes from the linear approximation in LSW which yields 
an error of the order of $1/S$ (which for $S=1$ could be large).
We will further discuss the INS result in subsection~\ref{sc:MC}.

Now we consider the bi-quadratic interaction between the NN along inter-$ab$ planes.      
The dependence of the total energy on angle between  the Ni magnetic moments reveals that 
if there is the bi-quadratic interaction in this compound.
The Ni$^{+2}$ ions in
\SNW{} are located in the lattice points of two tetragonal sub-lattices shifted by $(a/2,a/2,c/2)$.
To calculate the bi-quadratic coupling constants $B_{ij}$,
starting from a FM configuration, we compute the total energy of the magnetic configurations 
in which the direction of the magnetic moments in these
two sub-lattices are rotated by the angle $\theta$.   

\begin{figure}[tb!]
\begin{center}
\includegraphics[width=0.45\textwidth,angle= 0.00,clip]{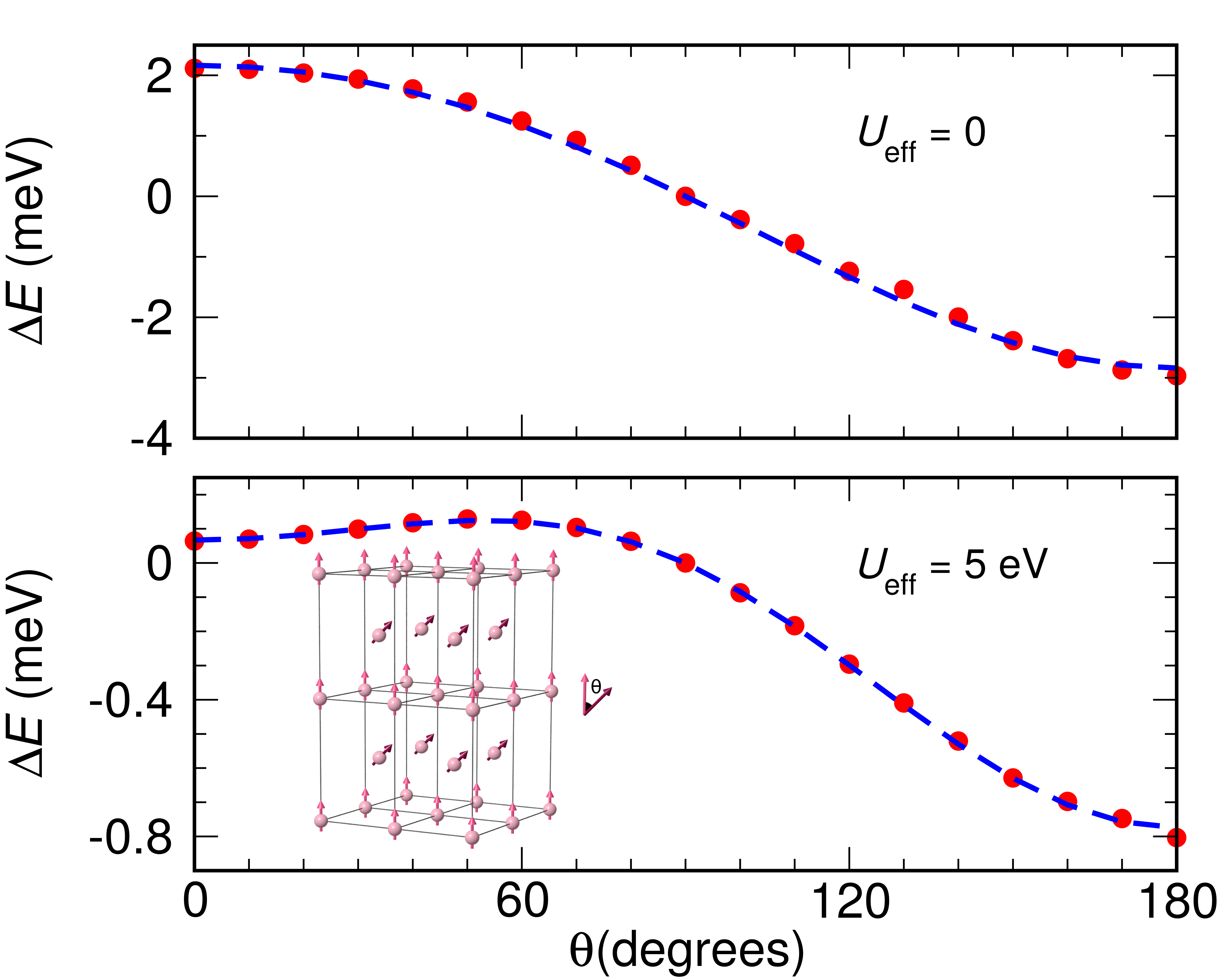}
\end{center}
\caption{(Color online)  
  Total energy versus rotational angle $\theta$ 
  (the angle between the magnetic moments of the two tetragonal sub-lattices).
  The reference of energy is set to $\theta=90\degree$. The dash line denotes the fit to the data
  using the function $f(x)=8(B\cos^2\theta-J_{1c}\cos\theta)$.
} 
\label{fig:bi}
\end{figure}


\begin{figure}[tb!]
\begin{center}
\includegraphics[width=0.45\textwidth,angle= 0.00,clip]{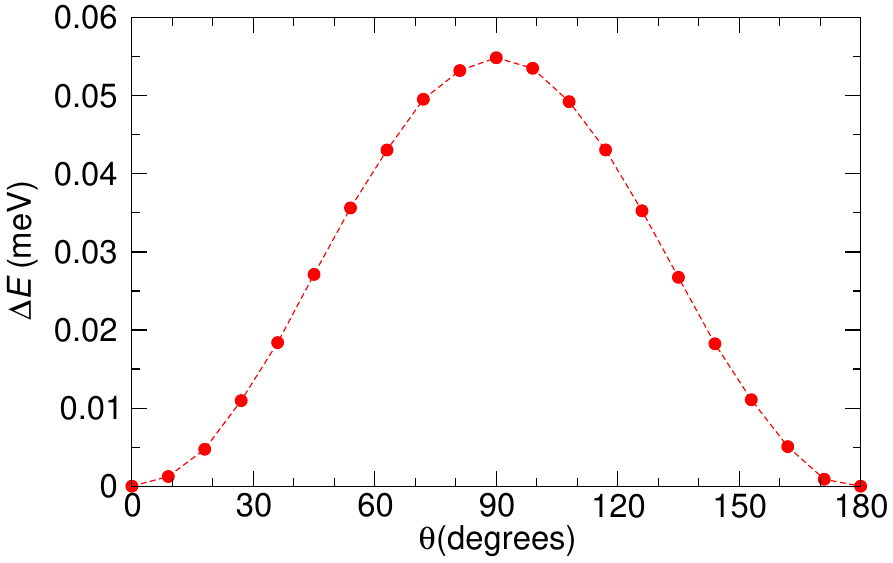}
\end{center}
\caption{
(Color online)  
  Total energy versus rotational angle $\theta$ respect to the lattice c-axis within GGA+$U$+SOC.
  The reference of energy is set to $\theta=0\degree$.
} 
\label{fig:sion}
\end{figure}

Fig.~\ref{fig:bi} presents the variation of  energy ($\Delta E=E(\theta)-E(\theta=90\degree)$) 
for GGA (i.e. $U_{\mathrm{eff}}=0$) and 
GGA+$U$ ($U_{\mathrm{eff}}=5$\,eV)  with the experimental structure.
 The $\Delta E$-$\theta$ curve can be well fitted by the function $f(x)=8(B\cos^2\theta-J_{1c}\cos\theta)$ which 
comes from the spin Hamiltonian containing only the inter-$ab$ plane NN Heisenberg
 and bi-quadratic interactions (the NNN interactions do not have any contribution in  $\Delta E$).
As a result, there is a bi-quadratic interaction in \SNW~in both of GGA and GGA+$U$.
 The bi-quadratic coupling constant is negative and  its value is, $B\approx-0.03$ and $~-0.04$\,meV for $U_{\mathrm{eff}}=0$ and $~5$\,eV, respectively.       
Similar results is obtained  if we use the optimized crystal structure with $U_{\mathrm{eff}}=4.72$\,eV.

Finally we investigate the single-ion anisotropy arising from the spin-orbit effect.
The single-ion term can be written as  $\Delta \sum _{i}  ({\hat{\bf n}}_i \cdot {\hat{\bf d}})^2$, 
where $\hat{\bf d}$ denotes the easy axis direction. We assume that $\hat{\bf d}$ is along lattice c-axis (see Fig.~\ref{fig:snwo}).
Using GGA+$U$+SOC with $U_{\mathrm{eff}}=5$, we calculated the total energy for the FM spin configuration as 
the angle $\theta$ (the angle between the magnetic moment direction and z-axis) varies from $0$ to $\pi$.
Fig.~\ref{fig:sion} shows the variation of $\Delta E$ versus $\theta$. 
This figure represent that the minimum energy is achieved at $\theta=0$. 
The value of $\Delta$ from this calculation is $\approx -0.05$\,meV, 
whose sign indicates that z-axis is indeed an easy axis.
We also checked that the total energy is independent of the azimuthal angle $\phi$.
It should be noted that the two-spin Ising anisotropy 
($J_{z} S_{i,z} S_{j,z} $) may have contribution to the energy difference curve in Fig.~\ref{fig:sion}.
Nevertheless, it is hard to separate its contribution since both single-ion and Ising anisotropy terms have the same angular dependence for uniform rotation of spin direction.
Due to this limitation we assign it totally to the single-ion anisotropy. 
\subsection{Monte Carlo Simulation}
\label{sc:MC}
For investigating the finite temperature behavior of the spin Hamiltonian 
we carry on classical MC simulations, using the coupling constants obtained by
  $U_{\rm{eff}}=0,~5,~6,~7$\,eV in experimental and $U_{\rm{eff}}=4.72$\,eV in optimized structure.
It is found that all the spin Hamiltonians, containing NN and NNN Heisenberg couplings 
(given in Table~\ref{tab:j}) and NN inter-$ab$ plane bi-quadratic terms ($B=J_{1c}$),
  in both experimental and optimized structures and the exchange couplings show a transition
 to a AFM-II ordering with the N\'eel  temperatures (\TNeel) given in  Table~\ref{tab:j}.
 Comparing the measured \TNeel$\approx 54$\,K~\cite{iwanaga, Blum2015}, we find that 
taking $U_\mathrm{eff}\approx 5$\,eV would be fine choice for the experimental structure.
 Moreover, in the optimized structure, the \TNeel~obtained by SCLR value $U_{\rm{eff}}=4.72$\,eV 
in a good agreement with the experimentally measured value. 
The SCLR method gives reasonable results for the compounds whose bonds have high  ionic character.
Using Critic2 code~\cite{critic2, critic}, we find the following valence state based on Bader charge analysis:
Sr$^{+1.60}_2$Ni$^{+1.39}$W$^{+3.00}$O$^{-1.26}_6$.
The charge analysis shows that the nature of Ni-O bonds 
in \SNW{} are predominantly ionic, which is the reason that SCLR works for this compound.         

The exchange constants obtained by INS, result to \TNeel$=34$\,K 
in the MC simulation which is $40\%$ less than  experimental value  \TNeel (54\,K).
This discrepancy, as already discussed, could be due to the using LSW for $S=1$ which underestimates 
the value of $J_2$.
\begin{figure}[tb!]
\begin{center}
\includegraphics[width=0.40\textwidth,angle= 0.00,clip]{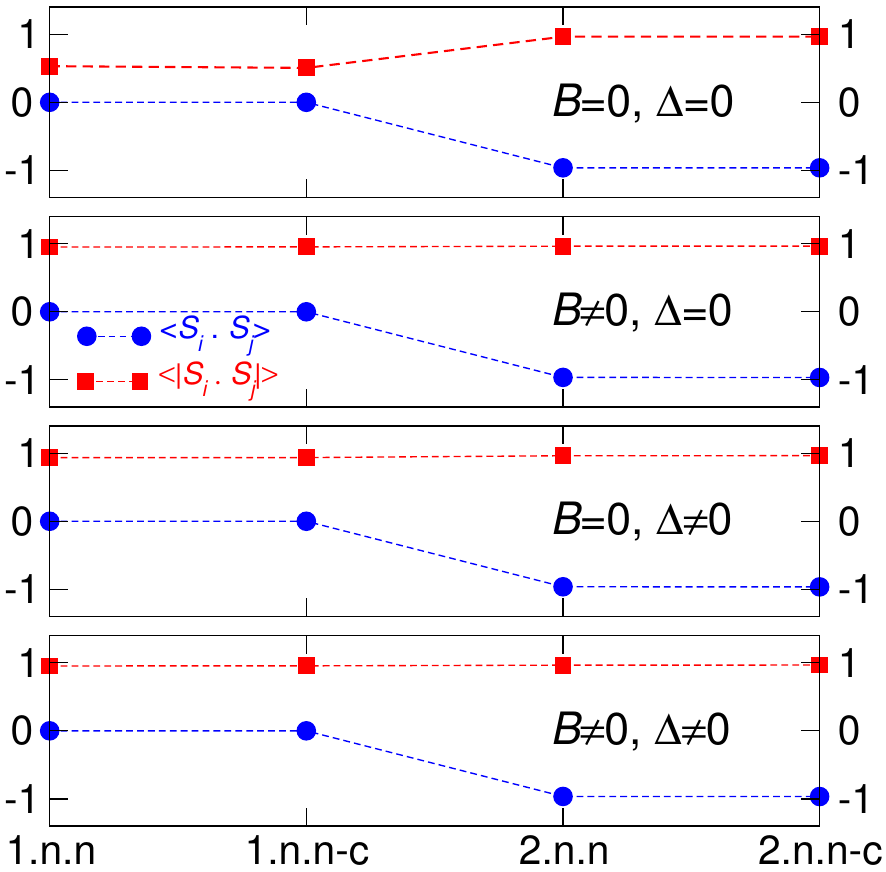}
\end{center}
\caption{(Color online) 
  Average (absolute) spin-spin correlation of intra- and inter- $ab$ plane NN and NNN 
  obtained by MC simulations at $T=4$\,K with the derived exchange parameters using $U_{\rm eff}=5$\,eV.}
\label{fig:sisj}
\end{figure}

It should be mentioned that the bi-quadratic and single-ion interactions do not have much effect on \TNeel, nevertheless
  we  will show in the following that they have an essential role in singling out a collinear spin configuration.  

To gain an insight into  the low temperature magnetic ordering in MC simulations, we calculate the average spin-spin correlation
at $T=4$\,K. Fig.~\ref{fig:sisj} represents the averages of the products of neighboring spins ($\langle S_{i}\cdot S_{j}\rangle$) and also 
their absolute values ($\langle |S_{i}\cdot S_{j}|\rangle$)  for the 
spin Hamiltonian given by the couplings obtained by $U_{\rm{eff}}=5$\,eV. 
As can be seen from this figure regardless of absence or presence of the bi-quadratic  and single-ion interactions, $\langle S_{i}\cdot S_{j}\rangle$ is 
 $\approx 0$ for both intra- and inter- $ab$ plane NN spins and   $\approx 1$ for all NNN spins. However, for the NN spins the value of
 $\langle |S_{i}\cdot S_{j}|\rangle$ is less than $1$ ($\approx 0.5$) in absence of 
bi-quadratic term or single-ion term  and about $1$  when including 
one or both of these interactions ($B\approx-0.04$, $\Delta\approx -0.05$\,eV).
 These calculations show that when  both $B$ and $\Delta$ are zero, 
the magnetic moments have freedom to rotate with 
respect to their nearest neighbors, however when either $B$ or $\Delta$ is turned on
 they loss their freedom and fix their directions parallel or anti-parallel to their neighbors,
 hence stabilizing the collinear configuration AFM-II. Indeed, the freedom of magnetic moments 
to rotate would give rise to  residual entropy at the low temperatures,
however the experimental results don't show such an entropy.\cite{Blum2015}

\begin{figure}[!t]
\begin{center}
\includegraphics[width=0.50\textwidth,angle= 0.00]{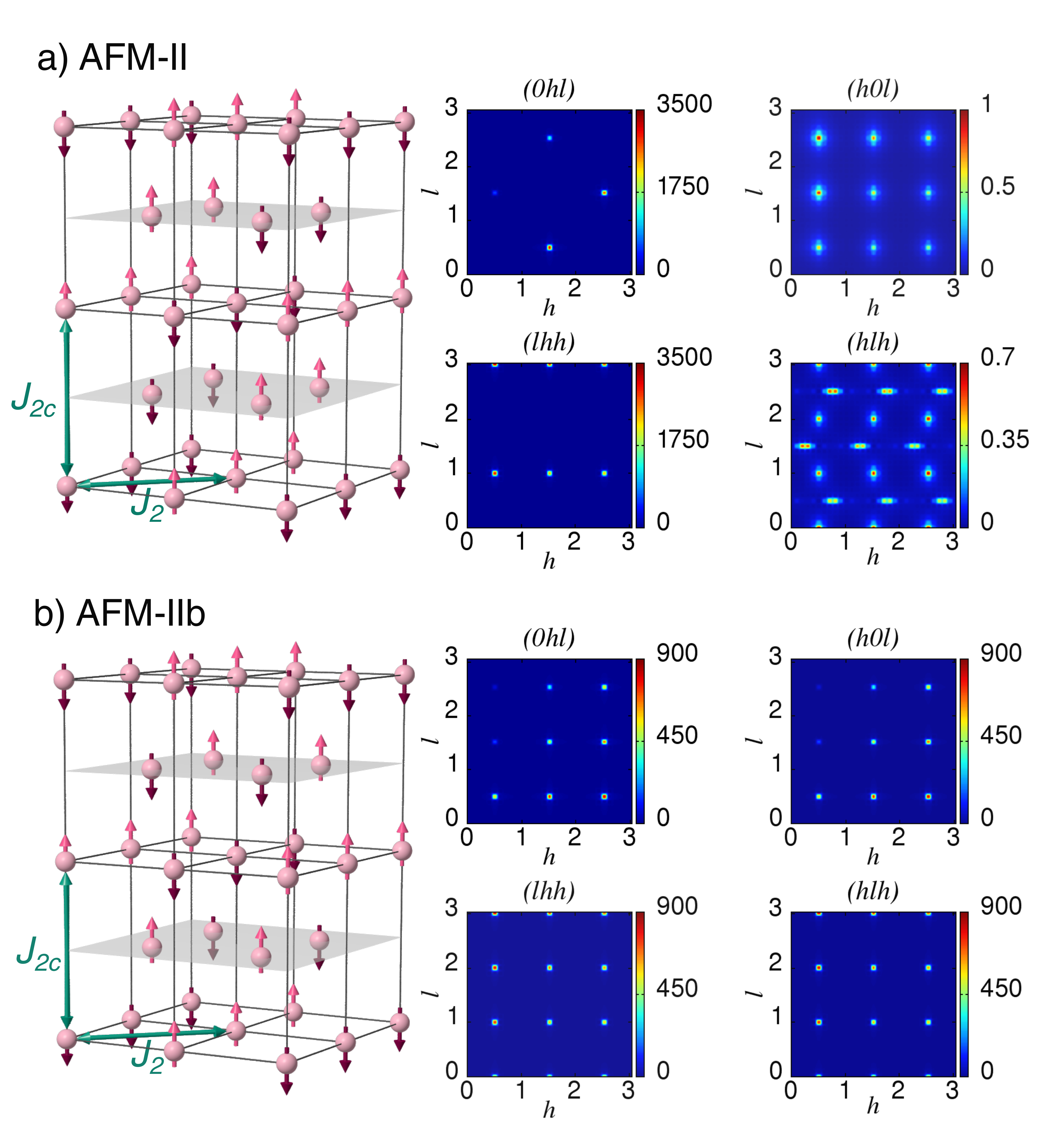}
\end{center}
\caption{(Color online)  A schematic of the type-II anti-ferromagnetic (AFM-II) and type-IIb anti-ferromagnetic (AFM-IIb) ordering and
their MC neutron structure factor, $S(\mathbf{q})$, in the ({\it 0hl}),  ({\it h0l}), ({\it lhh}) and 
 ({\it hlh}) planes at $T=4$\,K by including the Heisenberg, bi-quadratic and anisotropy terms in spin Hamiltonian which derived from GGA+$U$ ($U_{\rm eff}=5$\,eV).}
\label{fig:AF-II-IIb}
\end{figure}

Our MC simulations show that the ground state of \SNW{} is doubly degenerate. 
This can be verified by calculating the elastic neutron scattering structure function given by   
\begin{equation}
\begin{split}
S(\mathbf q)=&\sum_{i;j}\langle (\mathbf S_{i}-\frac{\mathbf S_{i}\boldsymbol{\cdot}\mathbf q}{\mathbf q\boldsymbol{\cdot}\mathbf q}\mathbf q) \boldsymbol{\cdot} 
(\mathbf S_{j}-\frac{\mathbf S_{j}\boldsymbol{\cdot}\mathbf q}{\mathbf q\boldsymbol{\cdot}\mathbf q}\mathbf q) \rangle  \\ 
&\exp[i\mathbf q \boldsymbol{\cdot}(\mathbf R_{i}-\mathbf R_{j})]
\end{split}
\end{equation}
Indeed, different MC runs end in two collinear spin configurations AFM-II and AFM-IIb illustrated 
in Fig.~\ref{fig:AF-II-IIb}.
The difference between these two configurations are the rotation of the $(0,0,2)$ planes 
(highlighted in gray) by 90\degree\ with respect to the $(0,0,1)$ planes along the $c$ direction. 
The right panels in Fig.~\ref{fig:AF-II-IIb} show the density plots
of $S(\mathbf q)$ for these two spin configurations. 
The main difference between the pattern of $S(\mathbf q)$ for these two spin configurations 
is the elimination of some Bragg peaks in AFM-II.  

While AFM-II and AFM-IIb are classically degenerate, it has been shown that in large $S$ limit 
the quantum effects lift the degeneracy of these two magnetic configurations 
in favor of AFM-II.\cite{quantum-frustration}

\section{CONCLUSIONS}
\label{conclusions}
In summary, we studied the magnetic interactions and thermodynamic properties of \SNW{}, 
using {\em ab initio} GGA and GGA+$U$ calculations and classical Monte Carlo simulation.
We found that interactions of the next nearest neighbors in the intra- and inter- $ab$ plane,
bi-quadratic interaction between inter-$ab$ plane nearest neighbors and the magnetic anisotropy along the $\hat{\bf z}$,
are the key players in determining the magnetic ordering of this compound.
Our results show that the classical ground state of \SNW{} has double degeneracy 
denoted by AFM-II and AFM-IIb. 
The elastic neutron scattering structure factors corresponding to these two magnetic 
configurations were calculated and presented as reliable theoretical references 
for experimental refinement of the true magnetic ground state of this compound by using 
neutron scattering experiments. 

\begin{acknowledgments}
N.R and H.A acknowledge the  support of   the National Elites Foundation and Iran National Science Foundation:INSF.
We acknowledge Hojjat Gholizadeh for his help to use POV-Ray.
\end{acknowledgments}

\bibliography{my_bibliography}

\section*{Appendix}
\subsection{Details about total energies of all our configurations}
\maketitle
The total energies for eight formula units without considering the nonmagnetic part, 
by Heisenberg Hamiltonian for ferromagnetic ordering can be written as:
\begin{equation}
E_{FM}= (32J_1+64J_{1c}+32J_2+16J_{2c}) S^2 
\end{equation}
and for considered AFM orderings would be as:
$$\setlength\arraycolsep{0.1em}
 \begin{array}{lccrcrcrcrcl}
E_{I}     &=& (& 32J_1&-&64J_{1c}&+ &32J_2&+ &16J_{2c}&) &S^2 \\
E_{II}    &=& (&  0J_1&+& 0J_{1c}&- &32J_2&- &16J_{2c}&) &S^2 \\
E_{III}   &=& (&-32J_1&+& 0J_{1c}&+ &32J_2&- &16J_{2c}&) &S^2 \\
E_{IV}    &=& (&-32J_1&+& 0J_{1c}&+ &32J_2&+ &16J_{2c}&) &S^2 \\
E_{V}     &=& (& 32J_1&+& 0J_{1c}&+ &32J_2&- &16J_{2c}&) &S^2 \\
E_{VI}    &=& (&  0J_1&+& 0J_{1c}&- &32J_2&+ &16J_{2c}&) &S^2 \\
E_{VII}   &=& (&-16J_1&+& 0J_{1c}&+ & 0J_2&+ &16J_{2c}&) &S^2 \\
E_{VIII}  &=& (&  0J_1&-&16J_{1c}&+ & 0J_2&+ &16J_{2c}&) &S^2 \\
E_{IX}    &=& (&  0J_1&+& 0J_{1c}&- &16J_2&+ &16J_{2c}&) &S^2 \\
E_{X}     &=& (& 16J_1&-&32J_{1c}&+ &16J_2&+ &16J_{2c}&) &S^2 \\
E_{XI}    &=& (& 32J_1&+& 0J_{1c}&+ &32J_2&+ & 0J_{2c}&) &S^2 
 \end{array}
$$

In table~{\ref{tab:dft-model}} we gather and  compare  the GGA+U/PAW total energy  with its counterpart Heisenberg Hamiltonian.
The mean absolute error of Heisenberg Hamiltonian energy respect to GGA+U/PAW total energy is about $0.07$\,meV/f.u. . 

\begin{table}[tbh!]
\caption{The total energy difference of different magnetic configurations (meV/f.u.) with respect to the ferromagnetic configuration within
GGA+U/PAW and Heisenberg Model. For GGA+U/PAW, we used optimized structure with $U_{\rm eff}=4.72$\,eV. For Heisenberg model, we used 
exchange parameters from optimized structure GGA+U/PAW with $U_{\rm eff}=4.72$\,eV. }
\label{tab:dft-model}
\begin{tabular}{l|cccc}
\hline
Method          &   GGA+U/PAW  &&&     Heisenberg Model     \\
\hline
AFM-I           &  -0.544    & & &  -0.368     \\
AFM-II          &  -16.264   & & &  -16.217      \\
AFM-III         &  -7.068    & & &  -6.954      \\
AFM-IV          &  -1.071    & & &  -1.147      \\
AFM-V           &  -5.848    & & &  -5.991      \\
AFM-VI          &  -10.440   & & &  -10.410     \\
AFM-VII         &  -5.759    & & &  -5.778      \\
AFM-VIII        &  -5.544    & & &  -5.584      \\
AFM-IX          &  -7.956    & & &  -7.974      \\
AFM-X           &  -2.817    & & &  -2.953      \\
AFM-XI          &  -3.051    & & &  -3.087      \\
\end{tabular}                     
\end{table}

\end{document}